\documentclass{elsart}

\usepackage{amssymb}
\usepackage{epsf,colordvi,color,amsbsy}
\usepackage[dvips]{graphicx}
\begin{document}

\begin{frontmatter}

\title{A general method for the determination of duration of solar cycle maxima}

\author[]{Stefano Sello \corauthref{}}

\corauth[]{stefano.sello@enel.it}

\address{Mathematical and Physical Models, Enel Research, Pisa - Italy}

\begin{abstract}

The use of different solar activity indices like sunspot numbers, sunspot areas, flare index, magnetic fields, etc., allows us to investigate the time evolution of some specific features of the solar activity and the underlying dynamo mechanism. One of the problems when using these activity indices for some statistical analyses is the reliable determination of the maximum phases of different solar cycles which are generally characterized by a multi-peaked structure due to the presence of the so-called \emph {Gnevyshev gap}.   
The main aim of this work is to propose a general method, without the introduction of ad hoc heuristic parameters, to determine the duration of a given solar cycle maximum phase through a long-term solar activity index like the Monthly Smoothed Sunspot Number (SSN). The resulting extended solar maxima allows us to include the multi-peaked structure of solar cycles and further the proposed method allows us to predict the solar maximum duration of the current solar cycle 24.

\end{abstract}
\end{frontmatter}

\section{Introduction}

The cyclic regeneration of the Sun large-scale magnetic field is the basic mechanism of all phenomena
collectively known as "solar activity". The solar magnetic cycle, due to the inductive action of fluid motions pervading the solar interior, it is the principal driver for all the visible surface manifestations, like the sunspots, flares, CME, etc.
While regular observations of sunspots go back to the early seventeenth century, and discovery of the sunspot cycle to 1843 by Schwabe, it is the  work of George Ellery Hale and collaborators that, in the twentieth century, demonstrated the magnetic nature of sunspots and of the solar activity cycle.
Sunspots appear when deep-seated toroidal flux ropes rise through the convective zone and emerge at the photosphere. 
Due to the long-term historical records of sunspots (groups, areas, etc.), the related indices are very valuable and useful to characterize the long-term evolution of solar cycles.
The concept of the sunspot number was developed by Rudolf Wolf of the Zurich observatory in the middle of the 19th century. The related sunspot series is called the Zurich or Wolf
sunspot number (WSN) series. The relative sunspot number Rz is defined as:

\begin{equation}
 Rz = k ~ (10G + N) 
\end{equation}

where: G is the number of sunspot groups, N is the number of individual sunspots in all groups visible on the solar disc and k denotes the individual correction factor, which compensates for differences in observational techniques and instruments used by different observers, and is used to
normalize different observations to each other.
The recording of the WSN series was terminated in Zurich in 1982. Since then, the sunspot number series is routinely updated as the International Sunspot Number, ISSN, provided by
the Solar Influences Data Analysis Center in Belgium, SIDC. The ISSN series is computed using mainly the same definition as WSN but it has a significant distinction from the WSN, i.e. it is based not on a single primary solar observation for each day, but instead uses a proper weighted average of more than 20 approved observers. The complexity of solar cycle dynamics is well documented in literature and it is well characterized by the complex set of quasi-periodic waves, both time localized and time persistent, as shown by different wavelet analyses (see: Sello 2000, 2003). For a comprehensive review on solar cycle features see: Usoskin, 2008. 
In this work we used the series of Monthly Smoothed Sunspot Numbers, SSN derived by the ISSN series, which is the most commonly used series to determine the global evolution or shape of a given solar cycle, both for amplitude and timing.
The recognition and characterization of a structured multi-peaked feature for the Schwabe solar cycle ($\approx 11$ year) maximum phase, date back to earlier works by Gnevyshev, 1963, 1967, mainly utilizing the time behavior of different heliographic latitudinal distributions of the  
outside eclipses intensity coronal line at $5303$ \AA. In particular, the author found the evidence for a bimodal structure of the coronal activity maximum of the $19^{th}$ cycle, where the first peak is located at the end of the increasing solar cycle and the second one, involving lower solar latitudes only, located near the early phase of the declining activity.
These first results, supported by successive works, Kopecky,  and Kuklin, 1969, Gnevyshev, 1977 including events in the photosphere and chromosphere, led the authors to conclude that there are two different processes or waves, partly superimposed in time, that may be responsible for the observed dual-peaked feature during the  maximum phase of the solar activity cycles. 
The existence of a structured maximum in solar activity was confirmed in successive works in which we analyzed the
processes involved in time behavior of large and complex active regions using data for sunspot areas, coronal magnetic energy maxima, global heliomagnetic fields and several solar-geophysical indices.
A later work by Feminella and Storini 1997, contains
accurate investigations of the maximum activity shape using an extended  set of various indices related to different solar layers: the relative sunspot number, as photospheric activity index; the $10.7$ cm radio flux, mainly as chromospheric activity index; the full solar disk $1-8$ \AA~ solar
X-ray background, as a density index for the quiet corona structure; and the monthly average of grouped chromospheric flares. Analyzing the time behavior of the above indices at different time-scales and for several cycles, the authors
confirm the significant existence of the multi-peaked structure of the maximum phases.
Moreover, the multi-structured maximum phases appear as a common characteristic of all the solar atmospheric layers considered. Another important aspect of the multi-peaked structure is its relation with the intensity and long-duration
or importance of the event analyzed, supporting that only particular energetic phenomena associated to the interaction between global and local magnetic fields are mainly involved in the onset of the solar maximum shape. In this context,
the gap associated to the bimodal feature (Gnevyshev gap), should occur during the space-time variability (reversal) of the general heliomagnetic field.
In this context it is very useful to be able to determine the extension or duration of a maximum phase of a given cycle, using a general criterion based on the different related dynamics that exists in the rising or descending phases and the maximum phase. Many proposals and definitions of a maximum phase have been documented in literature by different authors, but these include some heuristic or arbitrary element to define its duration. As an example, in a recent work by 
Kilcik and Ozguc, 2013, the authors define the duration of Solar Maxima 
of historical cycles (from Cycle 1 to Cycle 23) using the largest smoothed ISSN data as a first parameter and as a second parameter, the authors propose the $15\%$ of the maximum values of the largest monthly ISSN for each corresponding cycle and subtract them from the largest smoothed ISSN. Thus, the best SSN values that describe the smallest SSN of a maximum (SSNlimit) is given by:

\begin{equation}
 SSNlimit = A - 0.15 ~ B
\end{equation}

where: SSNlimit is the smallest SSN that describes both the beginning and end of maximum of a cycle, A is the largest smoothed monthly maximum SSN, and B is the monthly maximum SSN of the corresponding cycle. To reduce the fluctuations of monthly ISSN, the SSNlimit values are properly smoothed.
In order to avoid the introduction of heuristic and arbitrary parameters in the definition of solar cycle maxima duration, here we propose a general method based on the presence of inflection points in the smoothed cycle curve, determined by a proper computation of the second derivative and the selection of related inflection points, which are the signals of a global modification in the evolution of a cycle with a proper characterization of the (extended) maximum phase, which is different from the ascending and descending phases.

\section{Duration of a solar maximum phase}

The method used to define and to numerically compute the duration of a given solar maximum phase, is based on the presence of inflection points in the Monthly Smoothed Sunspot curve. An inflection point on a curve, $f(x)$, is defined on the basis of its second derivative value. As it is well known, an inflection point is a point on a curve at which the sign of the curvature (i.e., the concavity) changes. Moreover, a necessary condition for an inflection point, $x$, is that the corresponding second derivative is zero and a sufficient condition requires that the second derivatives have opposite signs in the neighborhood of the point on opposite sides:

\begin{equation}
 f"(x)=0 ~ and: ~ f"(x-\epsilon)>0 ~ if:  f"(x+\epsilon)<0 
\end{equation}

and viceversa, where: $f"(x)$ is the second derivative computed at $x$.

The basic idea of using this definition for determining the duration of a solar maximum phase is that an inflection point on the solar cycle curve reflects the true change in the evolution of the solar activity and, in particular, it is a clear sign of the end of the ascending, or a start of the descending, phases of cycle on both sides of a maximum phase. The further advantage of this method is that it is not necessary to introduce arbitrary heuristic parameters to define the extremes or boundaries of the maximum phase.
The algorithm used to evaluate the behavior of the second derivative on the Monthly Smoothed Sunspot curve (SSN values) is based on a standard numerically evaluation of the second derivative on a given point of a unidimensional curve and to select the proper inflection points that separate the maximum phase from the ascending and descending phases of the cycle. In fact, due to the presence of local fluctuations in the SSN curve, we need to select only the inflection points on both  sides of the maximum that correspond to a permanent or stable change in the behavior of the curve shape, i.e. avoiding some "local" inflection points that are not a definite sign of a change in the cycle dynamics.
Thus, this method determines an "extended" maximum phase of a given cycle, i.e. distinguishing the three main phases of a solar cycle: the ascending phase, the maximum phase and the descending phase. In this way we are able to include, in the above extended maximum phase, all the possible multi-peaked structures of the solar cycles. Another interesting feature of this definition is that the extreme values of a maximum phase, i.e. the start and the end values of the related SSN series, are not necessarily the same.

\section{Results and discussion}

The method described above to define and to numerically compute the duration of a given solar maximum phase, has been applied to the whole historical series of the Monthly Smoothed Sunspot Numbers (SSN) (cycles from 1 to 23) from the SIDC database. In order to define the start time and the end time of a given cycle we used the criterion based on the SSN minimum values.

The parameters computed for each cycle are:

1) tmax start = the time corresponding to the left side of a maximum;
 
2) tmax end = the time corresponding to the right side of a maximum;

3) SSN start = the SSN value at tmax start;

4) tmax = the time corresponding to the maximum SSN value;

5) SSN max = the SSN value at tmax;

6) SSN end = the SSN value at tmax end;

7) Dur. = the maximum phase duration in years.

The numerical values computed for all cycles from 1 to 23 are shown in table of Fig.1.
The duration of the maxima phases ranges from 2.99 yrs for cycle nr.8 to 5.83 yrs for cycles nr.4 and nr.6. The mean duration and its standard deviation for all 23 solar cycles are: $\mu=4.63$ yrs and $\sigma=0.80$ yrs, respectively.

 \begin{figure}[h!]
\resizebox{\hsize}{!}{\includegraphics{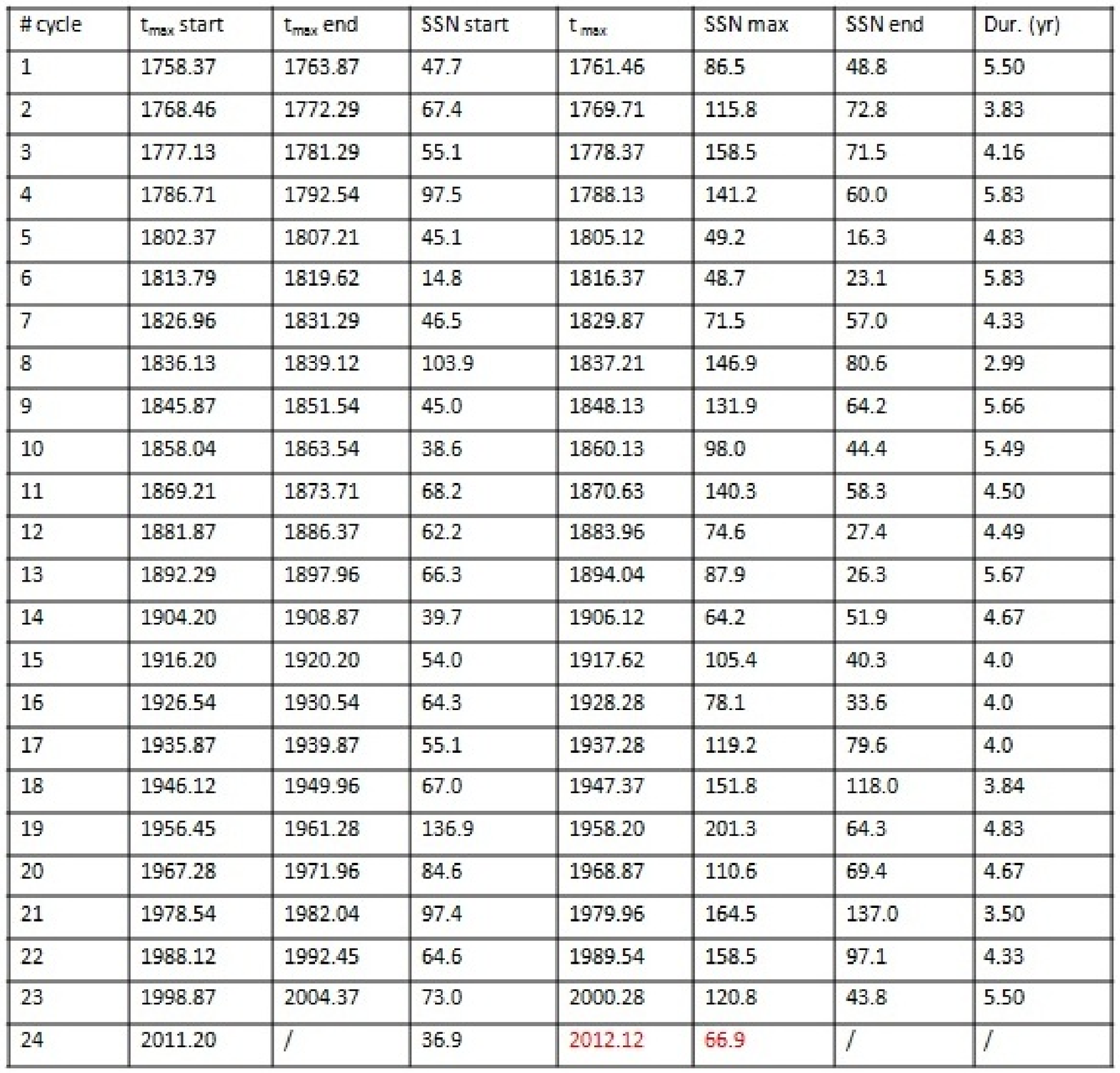}}
 \caption{Duration of the maximum phase of a given solar cycle using the second derivative method on the Monthly Smoothed Sunspot Numbers, SSN (SIDC data). Note that the SSN values computed at the start (SSN start) and at the end (SSN end) of the related maximum phase are not necessarily the same. The mean duration and its standard deviation for all 23 solar cycles are: $\mu=4.63$ yrs and $\sigma=0.80$ yrs, respectively. For cycle 24 we report the current provisional values.}
 \end{figure}

In order to use the duration of solar maximum phases as a possible statistically predictive tool, we tried to correlate this parameter with other useful parameters of the solar cycle. For example, we can compute the correlation existing between the duration and the SSN max, the SSN value at tmax. The result is shown in Fig. 2.

\begin{figure}[h!]
\resizebox{\hsize}{!}{\includegraphics{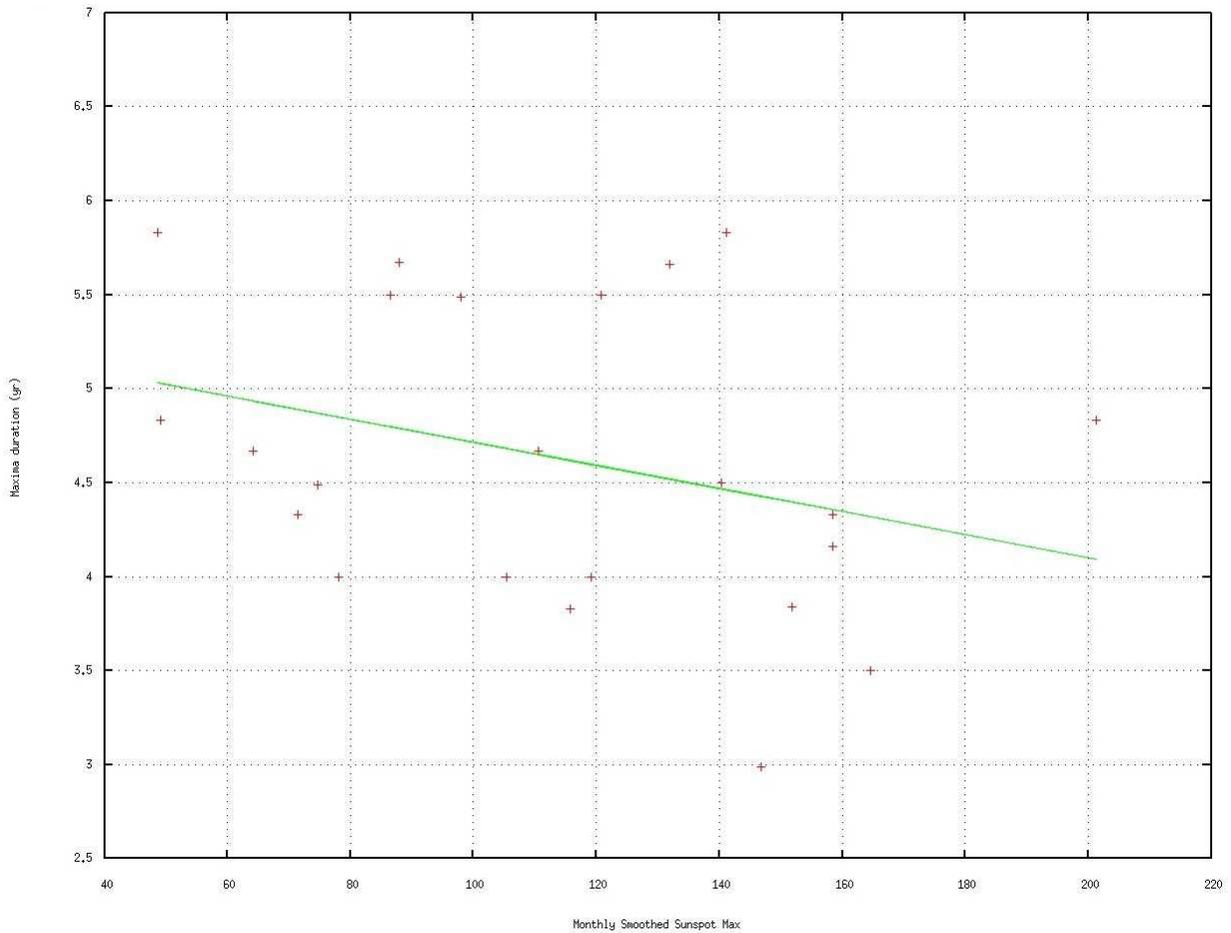}}
 \caption{Weak anti-correlation between Monthly Smoothed Sunspot Maxima, SSN, and the related Cycle Maxima Duration: R=-0.35.}
 \end{figure}

As we can see, there is a weak anti-correlation between Monthly Smoothed Sunspot Maxima and the related Cycle Maxima Duration with a correlation coefficient: $R=-0.35$. A similar result was obtained by Kilcik and Ozguc using their definition of solar maximum duration. Their average duration of solar maxima was: 2.91. This value appears quite low to well represent the real extension of the overall cycle maximum phases, if we consider that the average duration of the double peaked maxima with the presence of a Gnevyshev gap, is about 2 yrs.
However, if we correlate the Monthly Smoothed Sunspot Maxima and the corresponding ratio: Cycle Maxima Duration/SSN start, we obtain a stronger and more significant anti-correlation, as shown in Fig.3. Here the correlation coefficient is: $R=-0.64$ at $95\%$ significance level.

\begin{figure}[h!]
\resizebox{\hsize}{!}{\includegraphics{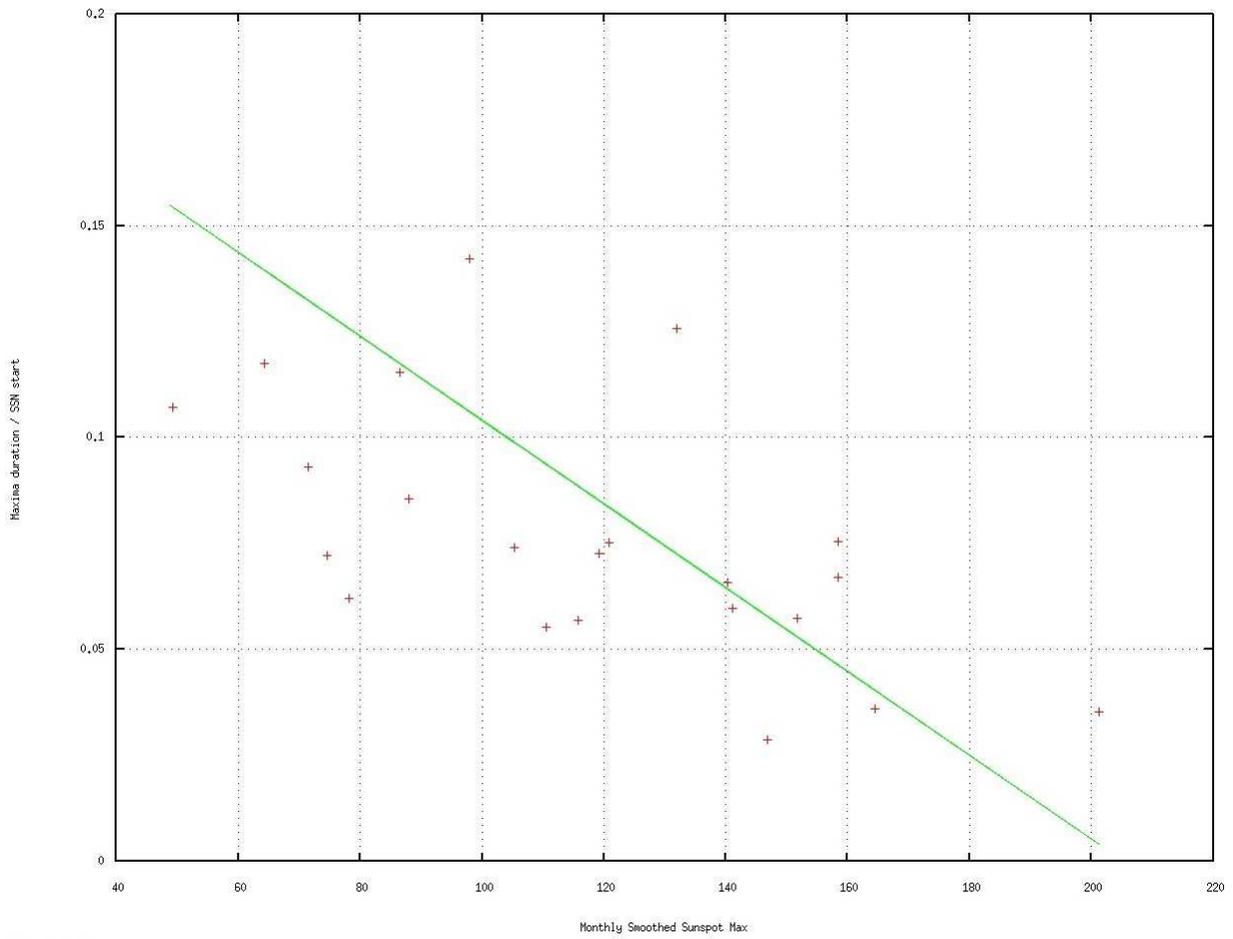}}
 \caption{Significant anti-correlation between Monthly Smoothed Sunspot Maxima, SSN, and the ratio: Maxima Duration/SSN start: R=-0.64.}
 \end{figure}

This last correlation allows us to use the related computed values to predict the solar maximum duration of solar cycles in progress.
For solar cycle 24 the known current provisional values are: $SSN start = 36.9$; $SSN max = 66.9$ (see table of Fig.1). With these values we can try to predict the duration of the maximum phase for solar cycle 24 using the above correlation. In this case, we find a statistically probable value in the interval: $3.4-4.5$ yrs, corresponding to a $t_{max} ~ end = 2014.6-2015.7$. This result indicates the highly probable onset of a multi-peaked structure for solar cycle 24 maximum.
Of course, the accuracy of this prediction cannot be very high due to the level of anti-correlation found, but it can be used, at least, as a first indication of a duration of the maximum phase for a given solar cycle in progress after its main peak.

With new data available coming from the next solar cycles it will be possible to improve the above analysis of the duration of solar maximum phases and to further check the reliability of all possible existing correlations with other cycle parameters. This step is essential to derive or to refine some statistical predictive tools.

\section{Conclusions}

In order to avoid the introduction of heuristic and arbitrary parameters in the definition of solar cycle maxima duration, here we proposed a general method based on the presence of inflection points in the smoothed cycle curve, determined by a proper computation of the second derivative and the selection of related inflection points, which are the signals of a global modification in the evolution of a cycle with a proper characterization of the (extended) maximum phase, which is different from the ascending and descending phases.

The main results obtained are:

1) The duration of the maxima phases ranges from 2.99 yrs for cycle nr.8 to 5.83 yrs for cycles nr.4 and nr.6. The mean duration and its standard deviation for all 23 solar cycles are: $\mu=4.63$ yrs and $\sigma=0.80$ yrs, respectively.

2) If we correlate the Monthly Smoothed Sunspot Maxima and the corresponding ratio: Cycle Maxima Duration/SSN start, we obtain a significant anti-correlation: $R=-0.64$ at $95\%$ significance level.

3) This correlation allows us to use the cycle computed values to predict the solar maximum duration of solar cycles in progress. For solar cycle 24 the known current provisional values predict a statistically probable value in the interval: $3.4-4.5$ yrs, corresponding to a $t_{max} ~ end = 2014.6-2015.7$. The onset of a multi-peaked structure for solar  cycle 24 maximum is thus highly probable. 

The accuracy of this prediction is related to the correlation coefficient of the anti-correlation 2) and it can be used, at least, as a first indication of a duration of the maximum phase for a given solar cycle in progress after its main peak.

Using new data coming from the next solar cycles it will be possible to improve the above analysis of the duration of solar maximum phases and to further check the existence of significant correlations with other cycle parameters, which is useful to refine some statistical predictive tools.

\section{References}

Feminella, F., and Storini, M., Large-scale dynamical phenomena during solar activity cycles, 1997, A\&A, 322, 311-319.

Gnevyshev, N.M., The corona and the 11-year cycle of solar activity, 1963, Sov. Astron., 7, 311-318.

Gnevyshev, N.M., On the 11-year cycle of solar activity, 1967, Sol. Phys., 1, 107-120.

Gnevyshev, N.M., Essential features of the 11-year solar cycle, 1977, Sol. Phys. 51, 175-183.

Kilcik A. and Ozguc A., One Possible Reason for Double-Peaked Maxima in Solar Cycles: Is a Second Maximum of Solar Cycle 24 Coming?, 2013, arXiv:1309.0731.
 
Kopecky, M., and Kuklin, G.V., A few notes on the sunspot activity in dependence on the phase of the 11-year cycle and on the heliographic latitude,  1969, Bull, Astr. Inst. Czech., 20, 22.

Sello, S., Wavelet entropy as a measure of solar cycle complexity, 2000, A\&A, 363, 311-315.

Sello, S. Wavelet entropy and the multi-peaked structure of solar cycle maximum, New Astronomy, Volume 8, Number 2, February 2003, pp. 105-117(13).

SIDC-team, World Data Center for the Sunspot Index, Royal Observatory of Belgium, Monthly Report on the International Sunspot Number, online catalogue of the sunspot index: http://www.sidc.be/sunspot-data/, 'year(s)-of-data'. 

Usoskin, I.G. A History of Solar Activity over Millennia, 2008
Living Rev. Solar Phys., 5, 3

\end{document}